\documentclass[10pt,twocolumn,letterpaper]{article}

\usepackage{cvpr}              
\usepackage{multirow}
\usepackage{url}
\usepackage{float}
\usepackage{comment}
\usepackage{booktabs}
\usepackage{booktabs}
\usepackage[dvipsnames]{xcolor}

\definecolor{cvprblue}{rgb}{0.21,0.49,0.74}
\usepackage[pagebackref,breaklinks,colorlinks,citecolor=cvprblue]{hyperref}

\newcommand{\case}[1]{\mbox{$\# #1$}}

\def\paperID{5} 
\def\confName{CVPR}
\def\confYear{2024}

\title{Event Camera Demosaicing via Swin Transformer and Pixel-focus Loss}

\author{Yunfan Lu, Yijie Xu, Wenzong Ma, Weiyu Guo, Hui Xiong\\
AI Thrust, Hong Kong University of Science and Technology (Guangzhou)\\
{\tt\small \{ylu066,yxu409,wma423,wguo395\}@connect.hkust-gz.edu.cn, xionghui@ust.hk}
}

\begin{document}
\maketitle
\begin{abstract}
Recent research has highlighted improvements in high-quality imaging guided by event cameras, with most of these efforts concentrating on the RGB domain.
However, these advancements frequently neglect the unique challenges introduced by the inherent flaws in the sensor design of event cameras in the RAW domain.
Specifically, this sensor design results in the partial loss of pixel values, posing new challenges for RAW domain processes like demosaicing.
The challenge intensifies as most research in the RAW domain is based on the premise that each pixel contains a value, making the straightforward adaptation of these methods to event camera demosaicing problematic.
To end this, we present a Swin-Transformer-based backbone and a pixel-focus loss function for demosaicing with missing pixel values in RAW domain processing.
Our core motivation is to refine a general and widely applicable foundational model from the RGB domain for RAW domain processing, thereby broadening the model's applicability within the entire imaging process.
Our method harnesses multi-scale processing and space-to-depth techniques to ensure efficiency and reduce computing complexity.
We also proposed the Pixel-focus Loss function for network fine-tuning to improve network convergence based on our discovery of a long-tailed distribution in training loss.
Our method has undergone validation on the MIPI Demosaic Challenge dataset, with subsequent analytical experimentation confirming its efficacy.
All code and trained models are released here: \href{https://github.com/yunfanLu/ev-demosaic}{https://github.com/yunfanLu/ev-demosaic.}
\end{abstract}

\section{Introduction}
\label{sec:intro}

\begin{figure}[t!]
\centering
\includegraphics[width=0.9\linewidth]{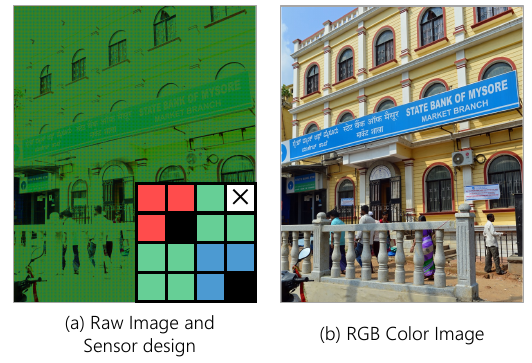}
\caption{\small Contemporary design of an actual event camera sensor (Hybridevs sensor), featuring red, green, and blue pixels for outputting RGB RAW signals. 
Black pixels in the lower right corner of the green and red areas are designated for event signal output, and white pixels do not emit any signals.
The demosaicing task aims to convert a RAW image with RGB signals and black holes (a) into a full-color image with three RGB channels (b).}
\label{fig:1-sensor}
\end{figure}

The event camera~\cite{gallego2020event,zheng2023deep,hybridevs2024mipi3}, with its low latency ($<100 \mu s$), high dynamic range ($>120 dB$), high temporal resolution ($>1000 fps$), and efficient power consumption, has garnered significant interest for enhancing computational imaging in applications, \eg,
video frame interpolation~\cite{tulyakov2021time,paikin2021efi,lu2023learning_rsc}, 
super-resolution~\cite{jing2021turning,lu2023learning}, 
deblurring~\cite{xu2021motion,sun2022event,jiang2020learning}, 
and high dynamic range~\cite{rebecq2019high,messikommer2022multi}.
These works are realized in the RGB domain, based on the premise that the camera sensor can simultaneously and seamlessly deliver RGB images and events.
These RGB images are obtained by RAW image processing approaches.
Specifically, RAW domain approaches convert RAW images, where each pixel contains only one type of color information with noise, into full-color images with all three RGB color information with high-quality~\cite{ignatov2022pynet}.
Filling in the missing color information is known as demosaicing~\cite{malvar2004high,hybridevs2024mipi3}, a core component of RAW domain image process.

However, the transformation of RAW to RGB faces significant challenges due to the existing event sensor chip design technology.
Specifically, event cameras produce RAW images where specific pixels are absent, as illustrated in Fig.~\ref{fig:1-sensor}.
These missing pixels emit event signals, not RGB signals, resulting in incomplete pixel values in the RAW output.
This absence of pixel values poses challenges for traditional RAW domain processing approaches, such as demosaicing, because the pixels emitting event signals cause the RAW image to lose \textbf{a quarter of} its red and blue color information.
To effectively address the challenge of enhancing downstream RGB images/videos with event guidance, it is imperative to transform incomplete RAW images into high-quality RGB full-color images without losing information.
Moreover, minimizing accumulated errors in this transformation process is essential for improving the quality of inputs for downstream tasks~\cite{ignatov2020replacing}.

RAW image process methods \cite{hirakawa2005adaptive,malvar2004high,su2006highly,zhang2005color,kokkinos2018deep,qian2022rethinking,a2021beyond,lu2022all} are predicated on traditional sensor technology, wherein each pixel is capable of capturing a color signal.
These methods can be divided into model-based~\cite{hirakawa2005adaptive,malvar2004high,su2006highly,zhang2005color} and learning-based methods~\cite{kokkinos2018deep,qian2022rethinking,a2021beyond}.
Learning-based approaches have attracted substantial interest due to the powerful fitting capabilities and robust generalizability of neural networks.
Inspired by the evolution of network architectures, these methods primarily aim to integrate and innovate neural network designs for RAW to RGB conversion mapping.
Many convolutional neural networks (CNNs)~\cite{kokkinos2018deep,qian2022rethinking,a2021beyond} were designed as the backbone for tasks such as demosaicing and denoising of RAW images.
For instance, PyNet~\cite{ignatov2020replacing} has designed a multi-scale, multi-resolution CNN network architecture for processing RAW into RGB.
More recently, benefits from the enhanced contextual modeling abilities and broader receptive fields from Vision Transformers (ViT)~\cite{dosovitskiy2020image}.
Many Transformer-based methods are proposed for image processing~\cite{shi2022video,lu2022video,wu2023video,zeng2023msfa,kim2023joint,xing2022residual}.
For example, RSTCANet~\cite{xing2022residual} employs the Swin-Transformer~\cite{liang2021swinir,geng2022rstt} to the demosaicing, incorporating global residual connections.
However, RSTCANet~\cite{xing2022residual} stacks Swin-Transformer layers~\cite{liang2021swinir,liu2021swin,liu2022video} without down-sample and multi-scale leads to higher computational complexity while failing to provide the network with a sufficiently large field of view.

Based on these considerations, we employ the Swin-Transformer-based backbone and a pixel-focus loss function for event camera demosaicing.
Our motivation is threefold: 
\textbf{(1) Scalability:} The Swin Transformer is a widely used and powerful foundational model in the RGB domain. Adapting it to the RAW domain could bridge foundational modeling across RAW and RGB imaging tasks. 
\textbf{(2) Efficiency:}  RAW domain methods are upstream of RGB domain processes and underpin all computer vision tasks. Therefore, RAW domain methods need to be sufficiently efficient to support downstream applications in the real world.
\textbf{(3) Training Effectivity:} A long-tail distribution of training loss was identified for the demosaicing task. Consequently, the pixel-focus loss was designed to facilitate a two-stage training process, enhancing the network's performance.

To achieve scalability, we refine the standard operators from the Swin-transformer~\cite{liang2021swinir,geng2022rstt} while avoiding customizations to enhance its portability. 
To ensure efficiency, we initially employ the space-to-depth~\cite{cheng2019learning} method to reduce the network's resolution and design a network structure akin to U-Net~\cite{ronneberger2015u}, achieving multi-scale and multi-resolution capabilities. 
This structure allows for a broader field of view with fewer layers. 
For effective training, we devised a two-stage loss function to fine-tune the network after completing the first training phase with \textit{Charbonnier} loss~\cite{lai2018fast}.

Our method underwent testing on the MIPI Demosaic Challenge dataset~\cite{hybridevs2024mipi3,mipi_2024} of the CVPR 2024 Workshop, demonstrating its applicability and performance. Subsequently, we conducted additional analytical experiments to evaluate its robustness and adaptability.
These tests solidified the method's effectiveness in various scenarios, clearly illustrating the superiority of our approach in addressing the intricacies of demosaicing in the RAW domain. Furthermore, we believe this work will inspire applications in the RAW domain and catalyze enhancements across multiple RAW-based tasks, fostering a new wave of innovation.

\section{Related Works}
\label{sec:formatting}

Modern digital cameras capture light, producing images with individual color channels (\eg, red, green, or blue) for each pixel~\cite{lyashenko2021modern}.
To compensate for the absence of color information, demosaicing is devised to reconstruct a full-color image from a single-channel RAW image~\cite{li2008image}.
In addition, owing to the wave-particle duality of light and the instability induced by dark currents in electronic devices, noise is a pervasive issue in the pixels of RAW images~\cite{brooks2019unprocessing}.
Consequently, the processes of denoising and demosaicing frequently occur concurrently.
Our paper focuses on the RAW domain processing of event cameras, prioritizing the demosaicing task due to its unique characteristics from the event camera sensor design.

\begin{figure*}
\centering
\includegraphics[width=1\linewidth]{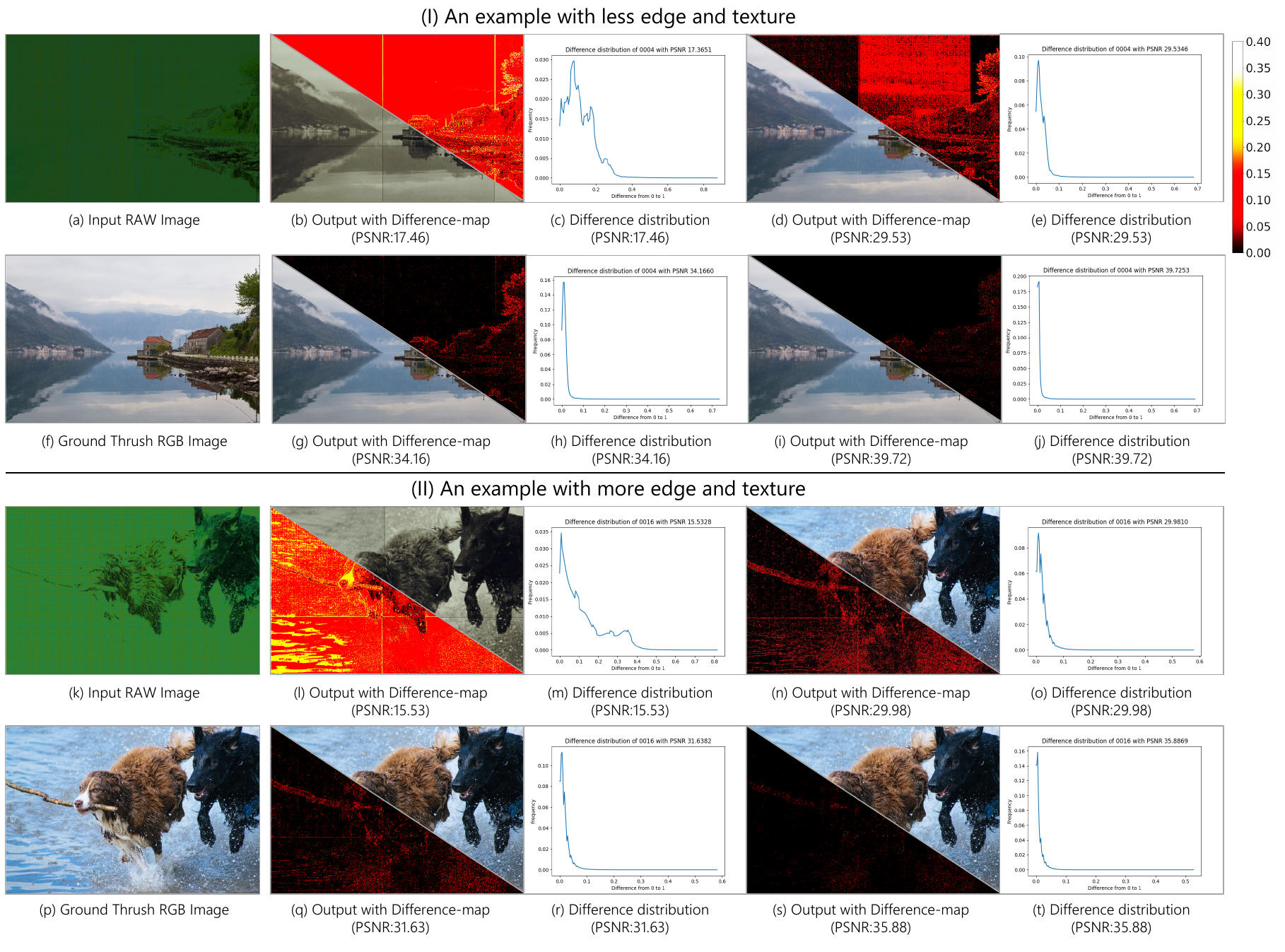}
\caption{\small Visual results of two images at different stages of training. Example (I) displays an image with less edge and texture, featuring extensive areas of sky and lake, while example (II) presents an image rich in edge and texture, including animal fur and splashing water.
For these two examples, four groups of reconstruction results are shown under varying PSNR values, along with different maps and difference distributions. 
Here, "difference" refers to the absolute value of discrepancies compared to the ground truth. 
As PSNR increases, the differences exhibit a long-tailed distribution.}
\label{fig:viz_losses}
\end{figure*}

\subsection{Camera RAW Image Demosaicing}
Demosaicing approaches can be categorized into two main groups: \textbf{(1)} model-based methodologies \cite{hirakawa2005adaptive,malvar2004high,su2006highly,zhang2005color}, which rely on mathematical models and spatial-spectral image priors for image reconstruction and \textbf{(2)} learning-based methodologies \cite{kokkinos2018deep,qian2022rethinking,a2021beyond,ignatov2020replacing,liu2020joint,xing2021end,syu2018learning,liu2021swin}, which leverage process mappings learned from extensive datasets of ground-truth images and corresponding mosaic counterparts. 
These techniques use different neural networks, \eg, CNNs and Transformers, to learn complex mappings between mosaic images and their corresponding full-color images. 
While CNN architectures have been widely used in learning-based demosaicing methods, they are limited in fixed-size receptive fields of convolution kernels and global context awareness compared with Transformer~\cite{xing2022residual}.
As a result, recent advancements in Transformer architectures, particularly the Swin-Transformer \cite{xing2022residual}, have shown promise in addressing these challenges and improving the performance of learning-based demosaicing approaches.

\begin{figure*}[ht!]
\centering
\includegraphics[width=\linewidth]{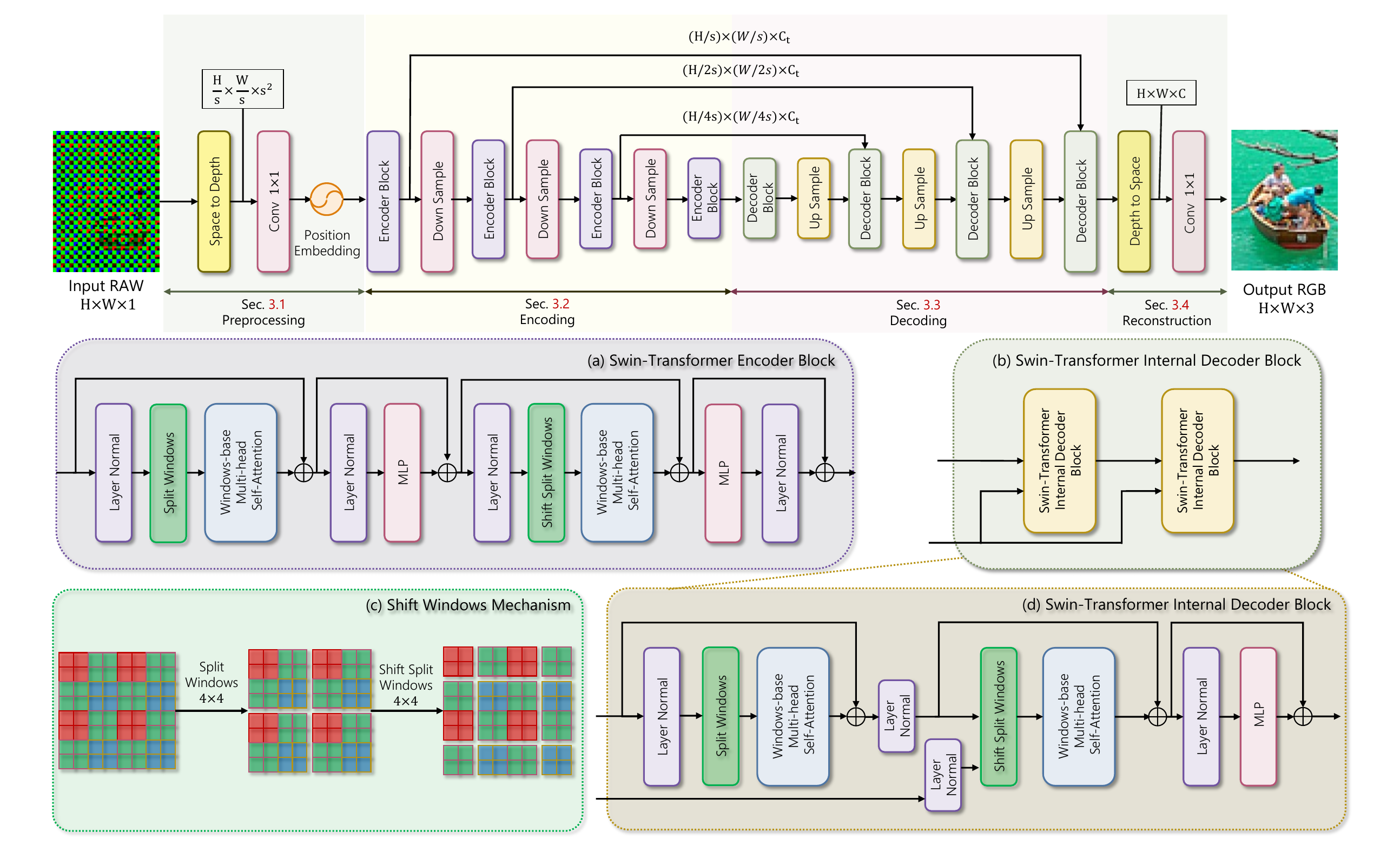}
\vspace{-15pt}
\caption{\small Overview of the event camera demosaicing method. The input RAW image is first preprocessed using space-to-depth and 1$\times$1 convolution operations. The encoder then extracts multi-scale features using Swin Transformer blocks with the shifted window mechanism. The decoder mirrors the encoder's structure and incorporates skip connections to recover spatial details. Finally, the reconstruction module generates the output RGB image. (a) Encoder block architecture. (b) Shifted window mechanism for cross-window interactions. (c) Decoder block architecture.}
\label{fig:framework}
\vspace{-5pt}
\end{figure*}

\subsection{Transformer-based Imaging Processes}
Transformer have been employed in many imaging processes, \eg, image/video super-resolution~\cite{yue2016image,lu2023learning}, deblurring ~\cite{zhang2022deep}.
Remarkably, Swin-Transformer \cite{li2023transformer,geng2022rstt,liang2021swinir} utilizes the shifted window mechanism to capture long-range dependencies in images, enabling effective aggregation of information from distant spatial locations, presenting impressive performance in vision tasks like super-resolution \cite{choi2023n,li2022hst},  video deblurring \cite{cao2022vdtr}, and video frame interpolation \cite{lu2022video,geng2022rstt}.
Besides, Swin-Transformer's hierarchical architecture partitions input images into smaller patches, which are then processed through multiple transformer blocks, facilitating the learning of both local and global features.
For example, ~\cite{liang2021swinir} utilizes a sequence of residual Swin-Transformer blocks for deep feature extraction, demonstrating leading-edge performance across various image super-resolution tasks.
Consequently, inspired by these successful works~\cite{geng2022rstt,liang2021swinir,cao2022vdtr} we also employ Swin-Transformer as a backbone to leverage the demosaicing.

\subsection{Image Reconstruction Loss Functions}
A series of works prefer Charbonnier Loss~\cite{lai2018fast} as the loss function to train their neural network for image reconstruction \cite{zhao2016loss,wang2022efficient}.
Nevertheless, areas characterized by numerous high-frequency details, \eg edges and textures, merit increased attention compared to low-frequency and smooth regions that are easily recoverable during the training process, as shown in Fig.~\ref{fig:viz_losses}.
In \cite{liu2020joint}, an adaptive-threshold edge loss is introduced to tackle this challenge, which adaptively adjusts the edge detection threshold for different image patches based on their edge density, allowing the model to focus more on regions with rich edge details during training.
However, the loss in \cite{liu2020joint} needs to divide the image into specific patches according to their edge density, demanding a series of complicated thresholds and cross-entropy loss calculations.
Consequently, we offer a facilitated approach called Pixel Focus Loss to optimize the model to capture subtle differences effectively.

\section{Methods}
This section presents the details of our event camera demosaicing method. 
As illustrated in Fig.~\ref{fig:framework}, our method leverages the strengths of the Swin-Transformer~\cite{liang2021swinir} and the U-Net architecture~\cite{ronneberger2015u} to take the RAW image with missing pixel values as input and aims to reconstruct a high-quality RGB image. 
Our framework consists of five key components: (1) preprocessing~\ref{sec:preprocessing} with a space-to-depth operation~\cite{sajjadi2018frame} and a $1\times 1$ convolution, (2) encoding~\ref{sec:encoding} with Swin-Transformer and shifted window mechanism, (3) decoding~\ref{sec:decoding} with mirrored encoding modules, (4) reconstruction~\ref{sec:reconstruction} with mirrored preprocessing modules, and (5) loss functions~\ref{sec:loss_function} with exquisite designs.

\begin{figure*}[t!]
\centering
\includegraphics[width=\linewidth]{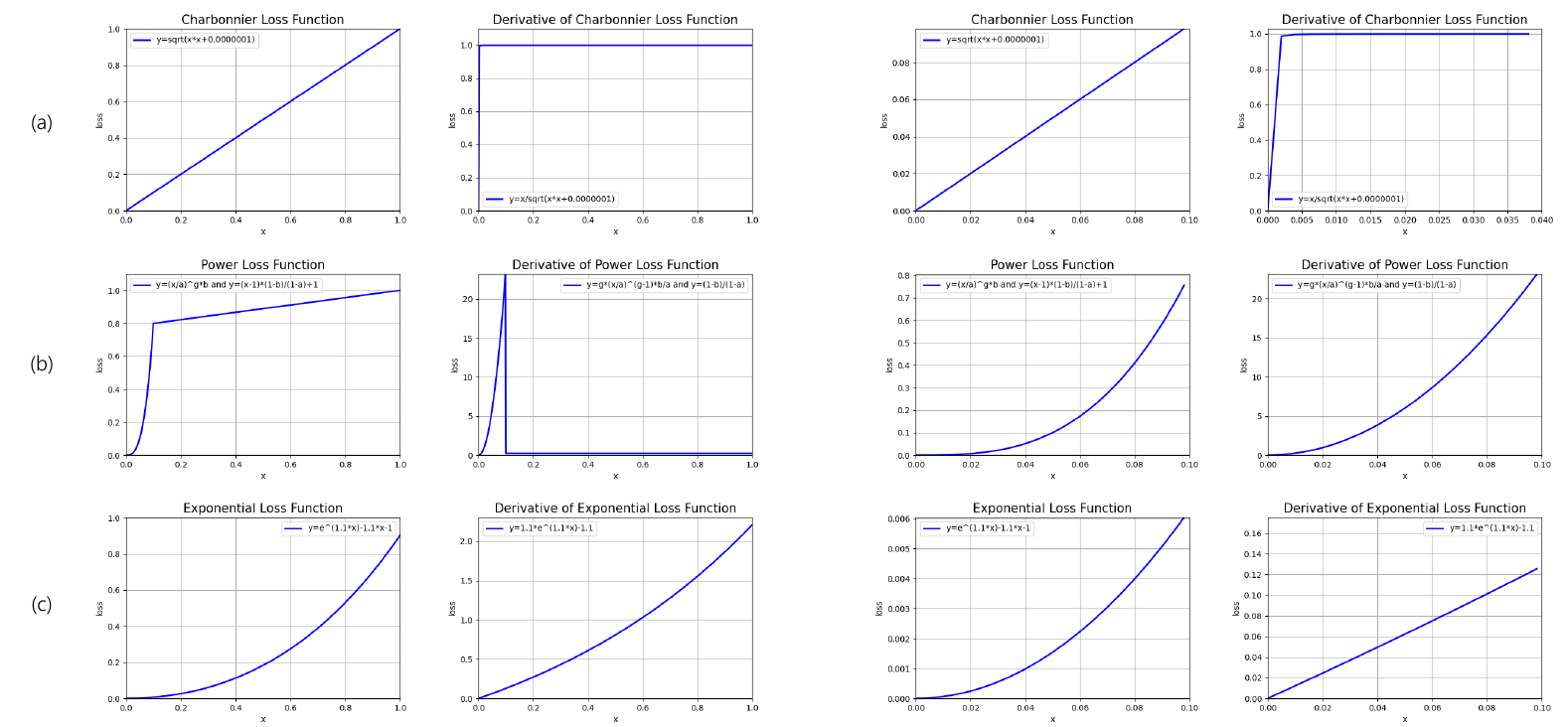}
\caption{\small Loss functions visualization. (a) (b) and (c) refer to Charbonnier and pixel-focus loss with the power and the exponential function, respectively. The line charts loss functions within the 0-1 range and their gradients. It also provides a magnified view of the 0 to 0.1 interval to observe the characteristics of different loss functions better when dealing with long-tail distributions.}
\label{fig:7-loss-function}
\end{figure*}

\subsection{Preprocessing\label{sec:preprocessing}}
The preprocessing module aims to transform the input RAW image into a suitable representation for the subsequent encoding stages while reducing the computational complexity.
Given the input RAW image $I_{RAW} \in \mathbb{R}^{H \times W \times 1}$, we apply a space-to-depth operation~\cite{cheng2019learning} with a factor of $s$ to reduce the spatial resolution to $(H/s) \times (W/s)$ and increase the channel dimension to $s^2$. 
This operation effectively reduces the model complexity, as the computational cost is linear with respect to the number of channels and quadratic concerning the spatial resolution~\cite{liang2021swinir}. 
Subsequently, a $1\times 1$ convolution is employed to generate the feature $F_0\in \mathbb{R}^{(H/s) \times (W/s) \times C}$ from the RAW image. 
To incorporate positional information, we add positional embeddings $E_{pos} \in \mathbb{R}^{(H/s) \times (W/s) \times C}$ to $F_0$. 
The positional embeddings~\cite{vaswani2017attention} are computed using a sinusoidal function:
\begin{equation}
\begin{cases}
E_{pos}(2 i)&=\sin \left(p / 10000^{2 i / C}\right) \\
E_{pos}(2 i+1)&=\cos \left(p / 10000^{2 i / C}\right),
\end{cases}
\end{equation}
where $p$ represents the position index and $i$ is the dimension index. The resulting preprocessed feature representation is obtained as $F_0^{'} = F_0 + E_{pos}$.

\subsection{Encoding\label{sec:encoding}}
The encoding module aims to extract multi-scale features and capture long-range dependencies. We adopt a U-Net-like architecture~\cite{ronneberger2015u} with the Swin-Transformer~\cite{liang2021swinir} as the backbone. 
The encoding module consists of four stages, each containing a Swin-Transformer block followed by a down-sample layer.
The Swin-Transformer block comprises a Layer Normalization (LN) layer~\cite{ba2016layer}, a Window-based Multi-head Self-Attention (W-MSA) module~\cite{liu2021swin}, and a Multi-Layer Perceptron (MLP).
The W-MSA module, illustrated in Fig.~\ref{fig:framework} (a), performs self-attention within local windows of varying sizes, allowing the model to capture multi-scale features and structural details at different granularities.
The Shifted Window mechanism, illustrated in Fig.~\ref{fig:framework} (c), is employed in alternating Swin-Transformer blocks to facilitate cross-window interactions and enhance the model's representational power.
After each Swin-Transformer block, a down-sample layer is applied to half the resolution of feature maps.
This multi-scale architecture enables the model to process information at multiple scales while progressively reducing the spatial resolution. 
Each downsampling operation reduces the computational complexity by a factor of 4 while quadrupling the receptive field, enabling the model to capture a large view field.

\begin{figure*}[t!]
\centering
\includegraphics[width=\linewidth]{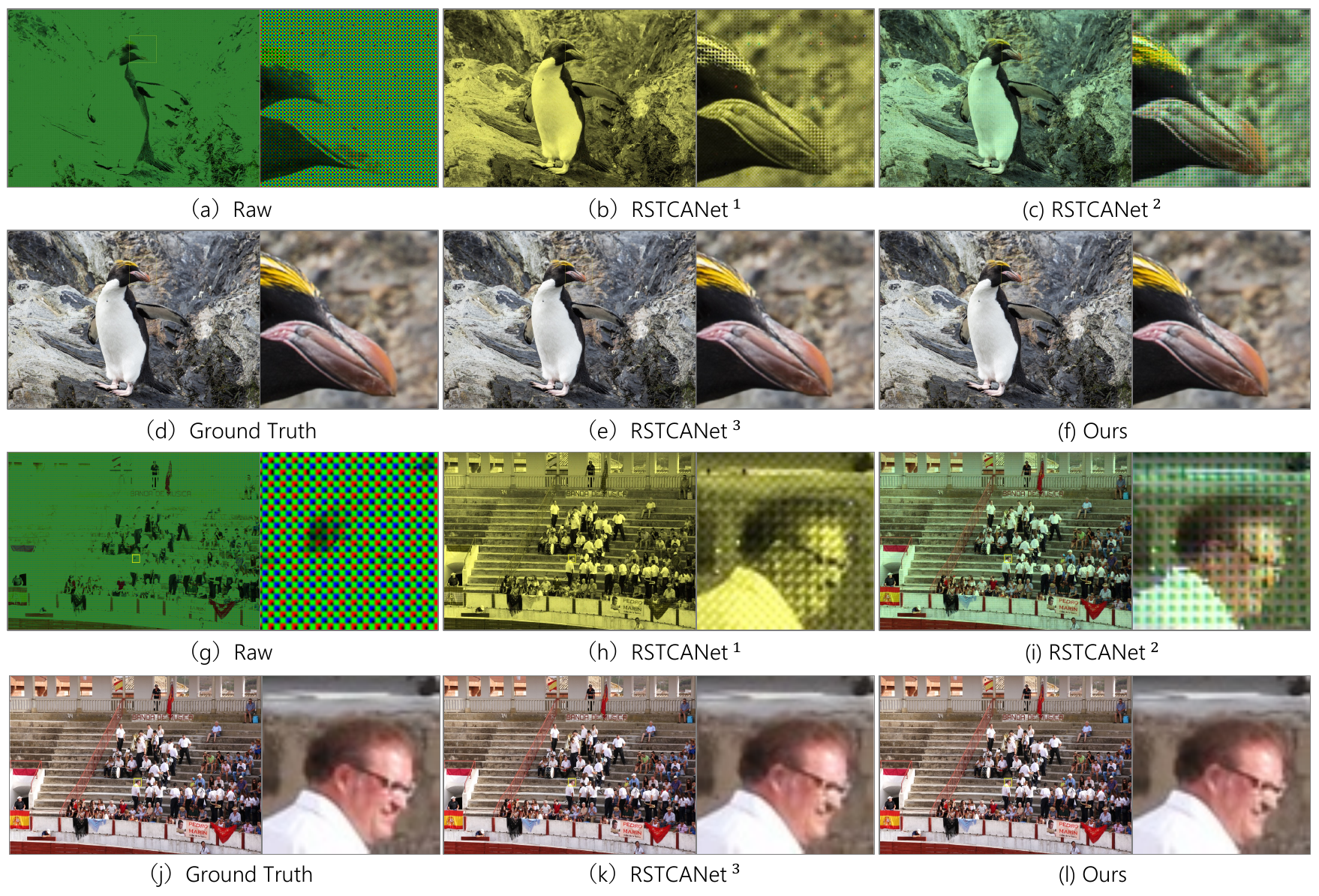}
\vspace{-20pt}
\caption{\small Visualized results of our method and compared method - RSTCANet~\cite{xing2022residual}. Comparison methods 1, 2, and 3, respectively represent the processing directly on the original RAW, processing after converting the original RAW into Bayer Pattern, and the results after fine-tuning RSTCANet~\cite{xing2022residual}.}
\label{fig:4-Comp}
\end{figure*}

\subsection{Decoding\label{sec:decoding}}
The decoding module aims to gradually upsample the feature maps and recover the spatial resolution of the output image. 
It follows a symmetric structure to the encoding module, consisting of four stages. 
Each decoding stage contains a decoder block, as depicted in Fig.~\ref{fig:framework} (b), followed by an up-sample layer. 
The decoder block comprises an LN layer, a W-MSA module, and an MLP, as illustrated in Fig.~\ref{fig:framework} (d). 
The W-MSA module in the decoder block operates similarly to its counterpart in the encoder block, capturing local dependencies within windows. 
An up-sample layer is employed after each decoder block to increase the spatial resolution.
Furthermore, skip connections are introduced between corresponding encoder and decoder stages to facilitate the flow of information and aid in recovering fine-grained details.
The multi-scale architecture of the decoding module enables the model to gradually refine the reconstructed image while incorporating features from different scales, leading to improved demosaicing performance.

\subsection{Reconstruction\label{sec:reconstruction}}
The reconstruction module aims to generate the final output RGB image from the upsampled feature maps produced by the decoding module. 
To achieve this, we apply a depth-to-space operation~\cite{shi2016real} to recover the resolution.
This step is crucial for maintaining image quality and minimizing distortions introduced during preprocessing. The depth-to-space operation rearranges the features and increases the spatial resolution by a factor of $s$, restoring the original spatial resolution of the input image. Finally, we employ a $1\times 1$ convolution to map the high-dimensional features to the three RGB channels with shape $H\times W\times 3$.

\subsection{Loss function\label{sec:loss_function}}

To train our demosaicing network, we employ a two-stage training approach. In the first stage, we use the Charbonnier loss~\cite{lai2018fast} for pre-training. The Charbonnier loss is a commonly used loss function in image processing as shown:
\begin{equation} 
\mathcal{L}_{Charbonnier} = \frac{1}{N} \sum\limits_{i=1}^{N} \sqrt{\left(I_{RGB}^{(i)} - I_{GT}^{(i)}\right)^2 + \epsilon^2},
\end{equation}
where $I_{RGB}^{(i)}$ and $I_{GT}^{(i)}$ represent the $i$-th reconstructed RGB image pixel and it's corresponding in the ground-truth image, respectively, $N$ is the total number pixels of images, and $\epsilon$ is a small constant (\eg, $1e-3$) added to improve the robustness of the loss function to outliers~\cite{lai2018fast}.
Evident from Fig.~\ref{fig:viz_losses} (I), the Charbonnier loss effectively reduces the difference with larger values during the pre-training stage.

\begin{figure*}[t!]
\centering
\includegraphics[width=0.95\linewidth]{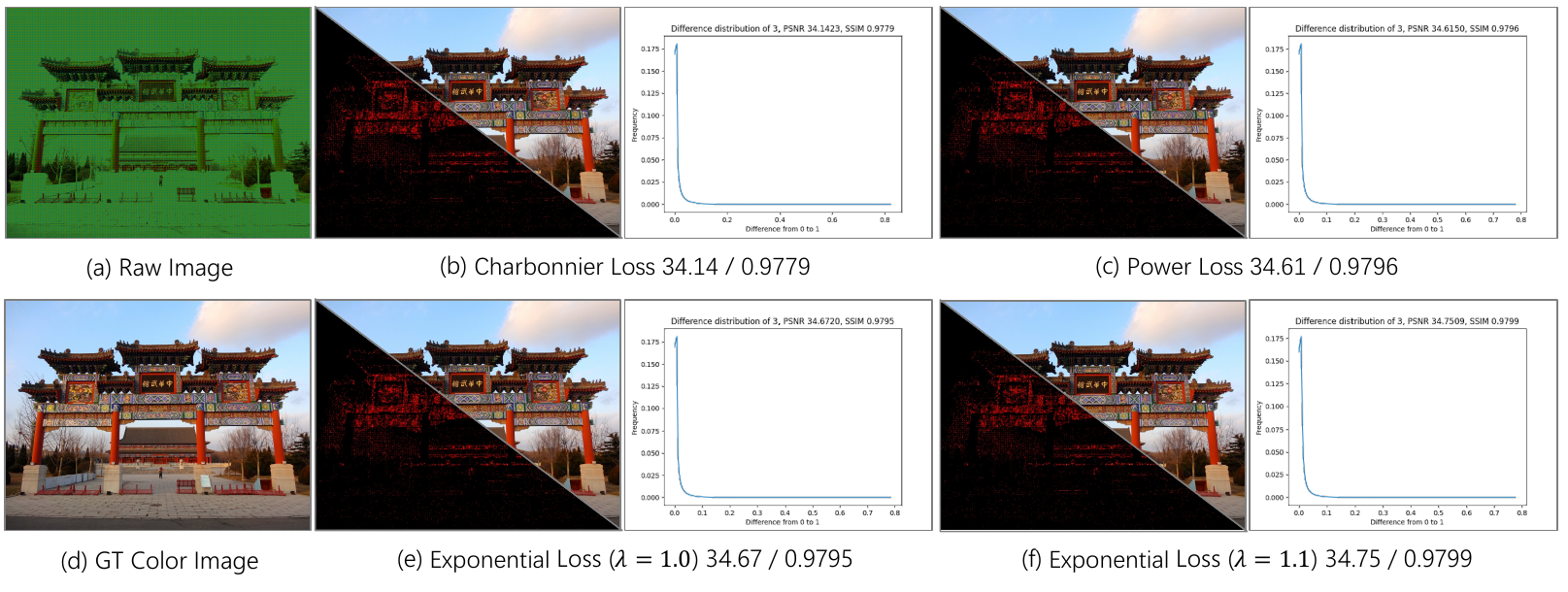}
\vspace{-5pt}
\caption{\small Visualization for output of fin-tuned models with different loss functions. The number values represent PSNR/SSIM.}
\label{fig:6-Different-Loss-Function-Release}
\end{figure*}

Upon closer examination of the Difference Distribution in Fig.~\ref{fig:viz_losses} (II), we observe that it comprises two main components: high frequency, \eg, edge areas, at low difference values and low frequency, \eg, smooth areas, at high difference values. 
This observation suggests that while the model efficiently learns to restore smooth blocks such as backgrounds, it has yet to handle edge areas fully.
This shortfall is primarily due to edge differences, highlighting the importance of increasing the gradient magnitude for edge.

In response to this need, we explored two forms of Pixel Focus Loss ($\mathcal{L}_{pf}$) to capture edge-related differences better. One version of the Pixel Focus Loss is described as a piecewise function:
\begin{equation}
\mathcal{L}^p_{pf} =
\begin{cases}
\left(d / a\right)^g \cdot b&,  0 < d < a \\
(d-1)(1-b) / (1-a) + 1&,  a \leq d \leq 1,\\
\end{cases}
\end{equation}
where $d$ represents the difference value, $a$ is a threshold parameter, and $b$ and $g$ are scaling factors that control the gradient magnitude. 
We also introduced another version of Pixel Focus Loss defined as:
\begin{equation}
\mathcal{L}_{pf}^{e}=e^{\lambda d}-\lambda d-1,
\end{equation}
where $\lambda$ serves as a hyperparameter, both versions address the issue, providing a comprehensive approach to fine-tuning the model's ability to capture edge areas. 
We utilize these two Pixel Focus Loss to fine-tune the pre-trained model obtained from the first stage, enhancing the demosaicing performance by emphasizing gradients for edge-related differences.
In the experiments section, we delve into the impact of different forms of Pixel Focus Loss and the effects of varying hyperparameters on the experimental results, providing a detailed analysis of our findings.
Combining the two-stage training approach, utilizing the Charbonnier loss for pre-training and the proposed Pixel Focus loss for fine-tuning, enables our network to learn high-quality RGB image reconstruction.
\begin{table}[t!]
\caption{\small Comparison of our method with the RSTCANet method~\cite{xing2022residual} on the MIPI-Challenge Demosaic dataset. RSTCANet 1, 2, and 3 denote different processing strategies.}
\label{tab:cmp-rstca}
\centering
\resizebox{1\linewidth}{!}{
\setlength{\tabcolsep}{0.029\linewidth}{
\begin{tabular}{c|c|c|cc}
\hline
Case     & Methods                        & Params ($M$)  & PSNR              & SSIIM  \\
\hline
\hline
\case{1} & RSTCANet$^1$                      & 7.19          & 13.2477           & 0.3590 \\
\case{2} & RSTCANet$^2$                    & 7.19          & 15.7721           & 0.4234 \\
\hline
\case{3} & RSTCANet$^3$                    & 7.19          & 36.2691           & 0.9659  \\
\hline
\case{4} & Our-\texttt{Small}              & 6.22          & 37.3272           & 0.9738 \\
\case{5} & Our-\texttt{Large}                & 9.40          & \textbf{37.4117}  & \textbf{0.9756} \\
\hline
\end{tabular}
}}
\end{table}

\begin{figure*}[t!]
\centering
\includegraphics[width=\linewidth]{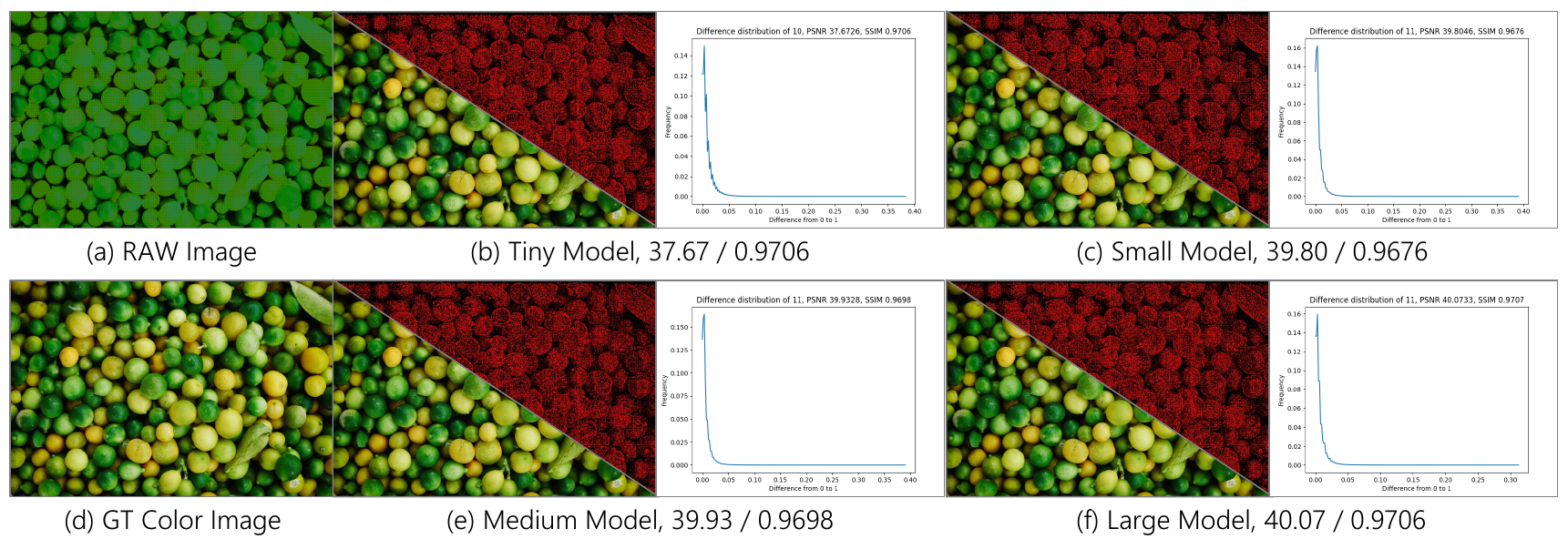}
\vspace{-15pt}
\caption{\small Visualization of results from model different size. The number values represent PSNR and SSIM, respectively.}
\vspace{-5pt}
\label{fig:5-Analytical-Model-Size}
\end{figure*}

\section{Experiments}

\noindent\textbf{Dataset:} 
Our experiments are based on the Demosaic for HybridEVs Camera dataset at MIPI-Challenge 2024~\cite{mipi_2024}, comprising 900 RAW-Color image pairs with around $2000 \times 1500$ resolution. 
In this dataset, 800 RAW-Color pairs are designated for training. 
50 color images are allocated for validation and another 50 for testing. 
Note that the validation and test sets were not released during the competition phase. 
Consequently, we adapted our approach by utilizing 760 images from the training set and designated 40 images for testing.
Within the validation set, 26 pairs of color images are available for local quantitative testing.

\noindent\textbf{Implementation Details:}
Our experiments were conducted using PyTorch~\cite{paszke2019pytorch} on a server with an Intel(R) Xeon(R) Platinum 8378A CPU and one NVIDIA A800 GPU.
The training batch size is one. 
Each training iteration employs random crop augmentation with patches sized at $640\times 640$.
In the first training phase, we utilized the Charbonnier Loss~\cite{lai2018fast}, training for 500 epochs with a learning rate starting from $1e-4$ and decreasing to $0$ following a cosine function.
In the second training phase, we applied the pixel-focus Loss, training for 200 epochs with a learning rate initiating from $1e-5$ and diminishing to $0$.
To achieve faster training speeds, we employed mixed precision techniques~\cite{micikevicius2017mixed} facilitated by PyTorch.

\noindent\textbf{Evaluation:} PSNR and SSIM~\cite{hore2010image} were utilized as quantitative evaluation metrics. 
Qualitative results are demonstrated through the visualization of difference maps and difference distributions.
Given the high performance of demosaicing methods, directly observing differences between images can be challenging; difference map visualization facilitates addressing this issue.

\subsection{Comparison Experiments:}

We benchmark our method against the publicly available RSTCANet~\cite{xing2022residual}, which is a pioneer in applying the Transformer architecture to demosaicing tasks. 
In contrast to our method, RSTCANet~\cite{xing2022residual} is tailored exclusively for demosaicing under the Bayer pattern and, consequently, is not inherently equipped to address scenarios with missing pixel values.
RSTCANet also falls short in addressing the missing values associated with defects in RAW images. 
Consequently, employing the RSTCANet method often results in images plagued with noise, as depicted in Fig.~\ref{fig:4-Comp}.

\begin{table}[t!]
\caption{\small Ablation for the model size. Params denote the model's parameters, measured in millions. Depth indicates the count of Transformer layers within each block}
\label{tab:model-size}
\centering
\resizebox{1\linewidth}{!}{
\setlength{\tabcolsep}{0.013\linewidth}{
\begin{tabular}{c|c|cc|cc}
\hline
Case & Model Size & Params ($M$) & Depth & PSNR    & SSIIM  \\
\hline
\hline
\case{1} & \texttt{Tiny}       & 4.62   & 2     & 36.3044 & 0.9696 \\
\case{2} & \texttt{Small}      & 6.22   & 4     & 37.3272 & 0.9738 \\
\case{3} & \texttt{Medium}     & 7.81   & 6     & 37.3798 & 0.9751 \\
\case{4} & \texttt{Large}      & 9.40   & 8     & \textbf{37.4117} & \textbf{0.9756} \\
\hline
\end{tabular}
}}
\end{table}

\begin{table}[t!]
\centering
\caption{\small Ablation for the Loss Function.}
\label{tab:loss-function}
\resizebox{1\linewidth}{!}{
\setlength{\tabcolsep}{0.053\linewidth}{
\begin{tabular}{c|l|cc}
\hline
Case & Loss Function             & PSNR    & SSIM   \\
\hline
\hline
\case{1} & w/o & 37.4117 & 0.9756 \\
\hline
\case{2} & $\mathcal{L}_{Charbonnier}$                    & 37.4402 & 0.9758 \\
\case{3} & $\mathcal{L}_{pf}^p$                         & 37.8164 & 0.9767 \\
\case{4} & $\mathcal{L}_{pf}^e$ with $\lambda=1$        & 37.8360 & 0.9765 \\
\case{5} & $\mathcal{L}_{pf}^e$ with $\lambda=1.1$      & \textbf{37.9656} & \textbf{0.9770} \\
\hline
\end{tabular}
}}
\end{table}

\subsection{Ablation and Analytical Experiments:}

\noindent\textbf{Ablation for the model depth and size:}
The ablation study indicates a progressive improvement in image reconstruction quality with increasing model depth, as shown in Tab.~\ref{tab:model-size} and Fig.~\ref{fig:5-Analytical-Model-Size}.
While the initial increase from a \texttt{Tiny} to a \texttt{Small} model shows a substantial rise in quality metrics, the growth tapers as the depth extends to \texttt{Medium} and \texttt{Large}.
This pattern suggests a diminishing return on enhancing PSNR and SSIM values with deeper networks, implying an optimal balance between depth and performance. 

\noindent\textbf{Ablation for the loss function:}
The Tab.~\ref{tab:loss-function} presented delineates an ablation study examining refining loss functions, particularly in mitigating the challenges posed by long-tail distributions encountered during the latter part of the initial training phase.
A secondary training phase spanning 200 epochs was implemented to counteract this issue, utilizing a suite of four variant loss functions. The benchmark was set using the Charbonnier loss function. Contrastive analyses were conducted with a power loss function and two exponential loss functions with $\lambda$ weights set at $1$ and $1.1$, as delineated in the second and fourth rows of Tab.~\ref{tab:loss-function}.
Notably, excessively high $\lambda$ values, such as $2$, have been observed to induce instability within the network. 
Evaluating the impact on PSNR and SSIM scores reveals that alterations to the loss function can significantly affect model performance.
The exponential loss functions, particularly with a $\lambda$ value of $1.1$, surpassed the performance of the baseline Charbonnier function. These insights imply that meticulous adjustment of the loss function parameters can effectively overcome optimization hurdles in the advanced stages of training, thereby improving the fidelity of the reconstructed images, as shown in Fig.~\ref{fig:6-Different-Loss-Function-Release}.

\section{Conclusion}
This paper employs Swin-Transformer and U-Net architecture tailored for the demosaicing task within the CVPR 2024 MIPI-Challenge. 
A pixel focus loss was designed for a two-stage training to facilitate efficient training. 
Our model demonstrates advantages over transformer-based methods for event camera demosaicing.
The model and the proposed loss function hold the potential to inspire future research in the field of demosaicing and beyond.

\noindent{\textbf{Acknowledgment:} 
This work was partially supported by Guangzhou-HKUST(GZ) Joint Funding 
Program (Grant No.2023A03J0008), Education Bureau of Guangzhou Municipality, Guangdong Science and Technology Department, and the 
Foshan HKUST Projects (FSUST21-FYTRI01A).
}

\bibliographystyle{ieeenat_fullname}
\bibliography{main}


\end{document}